# A Secure Multiple Elliptic Curves Digital Signature Algorithm for Blockchain

Wei Bi[*], Xiaoyun Jia, Maolin Zheng

Seele Tech Corporation, San Francisco, USA
weibi@seelenet.com

**Abstract:** Most cryptocurrency systems or systems based on blockchain technology are currently using the elliptic curves digital signature algorithm (ECDSA) on the secp256k1 curve, which is susceptible to backdoors implemented by the curve creator (secp256k1). The paper proposes a multiple elliptic curves digital signature algorithm (MECDSA), which allows not only for setting the number of elliptic curves according to practical security requirements, but also for editing the parameters of each elliptic curve. The performance analysis proves that the scheme is secure and efficient, and can avoid any backdoors implemented by curve creators. We suggest that the systems based on blockchain should operate in two elliptic curves considering the contradiction between security and efficiency.

## 1 Introduction

Digital signatures[1-2] are used to detect unauthorized modifications to data and to authenticate the identity of the signer. The recipient of signed message can use a digital signature as evidence to demonstrate to a third party that the signature was generated by the claimed signer. Digital signatures uses a combination of hash functions and public key cryptography. Firstly, the message digest is created using a hash function. The signature is created by encrypting the message digest using the signer's private key. Anyone can use the signer's public key and the same hash function to verify the received signature.

Until now, most cryptocurrency systems[3-5] or systems based on blockchain technology[6-7] have used in elliptic curves digital signature algorithm (ECDSA)[8-10], which is based on secp256k1[11]. Secp256k1 was almost never used before Bitcoin became popular, but it is now gaining in popularity due to several beneficial properties. Most commonly-used curves have random structure, but secp256k1 was constructed in a unique, non-random way which allows for especially efficient computation. As a result, it is often more than 30% faster than other curves if the implementation is sufficiently optimized. However, it cannot prevent the curve's creator from inserting any backdoors into the curve. So we propose a multiple elliptic curves digital signature algorithm

(MECDSA), which is more secure and can avoid any backdoors. In MECDSA scheme, the user cannot only choose the number of elliptic curves according to practical security requirements, but also can customize the curve by editing its parameters.

The multiple elliptic curves digital signature algorithm serves four purposes. Firstly, the signature proves that the owner of the private key, who is by implication the owner of the funds, has authorized the spending of the funds. Secondly, the proof of authorization is undeniable. Thirdly, the signature proves that the transaction has not and cannot be modified by anyone after it has been signed. Fourthly, the signature scheme can avoid any backdoors in the ECDSA curve.

In section 2, we summarize existing elliptic curve digital signature algorithm (ECDSA). In section 3, we propose a multiple elliptic curves digital signature algorithm (MECDSA). In section 4, we analyze the performance of MECDSA from its validity, security and efficiency.

## 2 Elliptic Curve Digital Signature Algorithm (ECDSA)

An elliptic curve E over a finite prime field Fp is defined by the short Weierstrass equation,
$$E: y^2 = x^3 + ax + b$$
where $a, b \in Fp$, $4a^3 + 27b^2 \neq 0 (mod p)$, and $p$ is a large prime. $P$ is a randomly selected elements on the elliptic curve E, called the base point, whose order $n$ is a large prime that makes $nP = O$, where $O$ is the zero element of the field $E(Fp)$, $n > 2^{160}, n > 4\sqrt{p}$.

A signer chooses a random number $d \in [1, n]$ as a private key, and computes public key $Q = dP$.

### 2.1 ECDSA Signature Generation

A signer signs the message $m$ according to the following steps:
Step 1: Compute $e = H(m)$, where $H$ is the secure hash algorithm.
Step 2: Select a random integer $k$ from $[1, n-1]$, and compute $kP = (x, y)$.
Step 3: Compute $r = x (mod n)$. If $r = 0$, go back to step 2 and reselect $k$.
Step 4: Compute $s = k^{-1}(e + dr)(mod n)$. If $s = 0$, go back to step 2 and reselect $k$.

Then the signature on the message m is the pair $(r, s)$.

### 2.2 ECDSA Signature Verification

Anyone can check to see if the signature $(r, s)$ of the message $m$ is valid by completing the following steps:
Step 1: Verify that $r$ and $s$ are integers in $[1, n-1]$. If not, the signature is invalid.
Step 2: Compute $e = H(m)$, where $H$ is a same hash function used in signature generation.
Step 3: Compute $w = s^{-1}(mod n)$.

Step 4: Compute $u = ew(mod\, n), v = rw(mod\, n)$.
Step 5: Compute $R = uP + vQ = (x, y)$.
The signature is valid if $r = x(mod\, n)$, or otherwise it is invalid.

The signer can obviously operate the ECDSA $t$ times (t-ECDSA), and get the signature $(r_1, s_1, r_2, s_2, ..., r_t, s_t,)$ in $t$ elliptic curves, but this will make the length of the signature long. To fix this, we present a secure multiple elliptic curves digital signature algorithm (MECDSA) in the next section, which can reduce the length of signature.

## 3  Multiple Elliptic Curves Digital Signature Algorithm

In this section, we propose a secure multiple elliptic curves digital signature algorithm (MECDSA), which can avoid any backdoors in the curve used by secp256k1. In this signature scheme, the user can not only choose the number of elliptic curves according to practical security requirements, but also can choose the curve by changing its parameters.

The parameters of MECDSA are same as ECDSA. There are $t$ elliptic curves $E_i: y^2 = x^3 + a_i x + b_i$ over finite prime field $Fp_i$, $P_i$ is the base point whose order is $n_i$, where $i = 1, 2, ..., t$.

In MECDSA, the signer chooses the number of curve and the parameters of the curves. After this, the signer chooses t random numbers $d_i \in [1, n_i]$ as a private key, and computes the public key $Q_i = d_i P_i$, where $i = 1, 2, ..., t$.

### 3.1  MECDSA Signature Generation

A signer signs the message m according to the following steps (fig. 1):
Step 1: Compute $e = H(m)$, where $H$ is the secure hash algorithm.
Step 2: Select random integer $k_i \in [1, n_i - 1]$, computer $k_i P_i = (x_{i1}, y_{i1})$, where $i = 1, 2, ..., t$.
Step 3: Compute $r_i = x_i (mod\, n_i)$. If $r_i = 0$, go back to step 2 and reselect $k_i$, where $i = 1, 2, ..., t$.
Step 4: Compute $r = r_1 + r_2 + \cdots + r_t$. If $r = 0(mod\, n_i)$, return step 2, where $i = 1, 2, ..., t$.
Step 5: Compute $s_i = k_i^{-1}[e + d_i r](mod\, n_i)$. If $s_i = 0$, go back to step 2 and reselect $k_i$, where $i = 1, 2, ..., t$.
Then $(r, s_1, s_2, ..., s_t)$ is taken as the signature of the message $m$.

### 3.2  MECDSA Signature Verification

Anyone can check whether or not the signature $(r, s_1, s_2, ..., s_t)$ of $m$ is valid by completing the following steps:
Step 1: Verify that $r$ and $s_i$ are integers in $[t, n_1 + n_2 + \cdots + n_t - t]$ and $[1, n_i - 1]$ respectively, where $i = 1, 2, ..., t$. If not, refuse the signature
Step 2: Compute $e = H(m)$, where $H$ is a same hash function used in signature generation.

Step 3: Compute $w_i = s_i^{-1}(mod\, n_i)$, where $i = 1,2,...,t$.
Step 4: Compute $u_i = ew_i(mod\, n_i), v_i = rw_i(mod\, n_i)$, where $i = 1,2,...,t$.
Step 5: Compute $R_i = u_i P_i + v_i Q_i = (x_i, y_i)$, where $i = 1,2,...,t$.
Step 6: Compute $r'_i = x_i(mod\, n_i)$, where $i = 1,2,...,t$.
If $r = r'_1 + r'_2 + \cdots r'_t$, accept the signature, otherwise, refuse the signature.

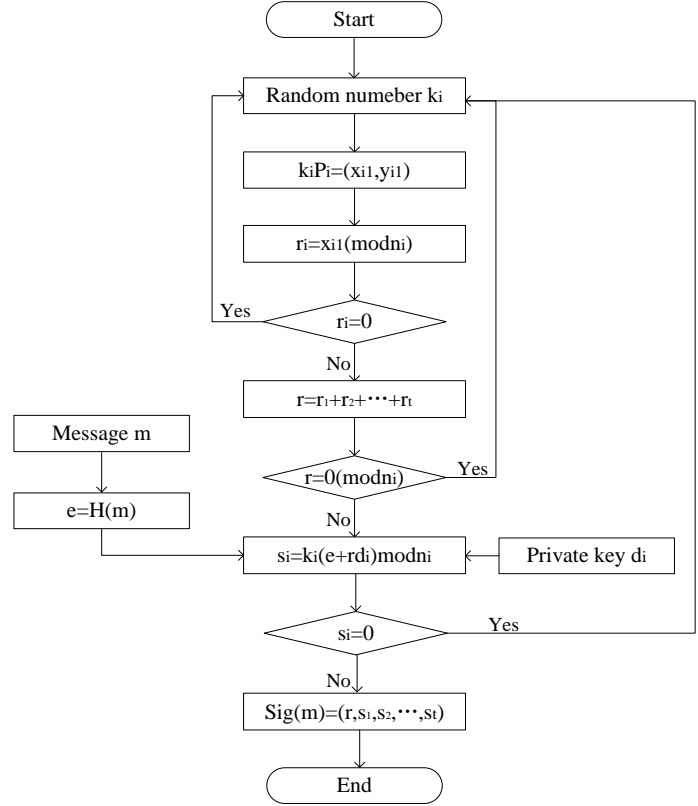

**Fig.1** MECDSA Signature Algorithm

## 4 Performance Analysis

In the section, we analyze the performance of MECDSA from different perspectives, including validity, security and efficiency.

### 4.1 Validity Analysis

According MECDSA signature generation and signature verification algorithm in MECDSA, we can get:
$$u_i = ew_i(mod\, n_i) = es_i^{-1}(mod\, n_i) = ek_i[e + d_i r]^{-1}(mod\, n_i)$$

$$v_i = rw_i(mod\,n_i) = rs_i^{-1}(mod\,n_i) = rk_i[e + d_i r]^{-1}(mod\,n_i)$$

So
$$\begin{aligned}
R_i &= u_i P_i + v_i Q_i \\
&= [ek_i(e + d_i r)^{-1}(mod\,n_i)]P_i + [rk_i(e + d_i r)^{-1}(mod\,n_i)]Q_i \\
&= [ek_i(e + d_i r)^{-1}(mod\,n_i)]P_i + d_i[rk_i(e + d_i r)^{-1}(mod\,n_i)]P_i \\
&= [k_i(e + d_i r)^{-1}(e + d_i r)(mod\,n_i)]P_i \\
&= (k_i(mod\,n_i))P_i \\
&= k_i P_i \\
&= (x_i, y_i)
\end{aligned}$$

So $r'_i = x_i mod\,n_i = r_i$ and $r = r'_1 + r'_2 + \cdots r'_t = r_1 + r_2 + \cdots r_t$, where $i = 1,2,\ldots,t$. Therefore, we prove the correctness of MECDSA.

### 4.2 Security Analysis

The security of cryptosystems using elliptic curves hinges on the intractability of discrete logarithm problem in the algebraic system. Historically, this problem has received considerable attention from leading mathematicians and cryptologists around the world[12-14]. Unlike the case of the discrete logarithm problem in finite fields, there is no subexponential-time algorithm known for the elliptic curve discrete logarithm problem (ECDLP). The best algorithm known to data takes exponential time.

The security of the proposed MECDSA is based on the difficulty of solving the ECDLP, and is mainly faced with the following attacks. MECDSA security is presented by the following analysis of attacks:

(1) The attacker wishes to obtain private key $d_1, d_2, \ldots, d_t$ using all information that is available from the scheme. In this case, the attacker can compute $r_1, r_2, \ldots r_t$ according to the known public key $Q_1, Q_2, \ldots, Q_t$, signature $(r, s_1, s_2, \ldots, s_t)$ and message m, and can further compute $K_i$ where $K_i = k_i P_i$. If an attacker wanted to obtain the private key, he would need to solved $Q_i = d_i P_i$ or $K_i = k_i P_i$ respectively for $d_i$ and $k_i$, which is clearly infeasible because the difficulty of solving the ECDLP, where $i = 1,2,\ldots,t$.

(2) The attacker wishes to obtain the private key $d_{i1}, d_{i2}, \ldots, d_{it_1}$ from a partial private key $d_{j1}, d_{j2}, \ldots, d_{jt_2}$, where $t_1 + t_2 = t$. When the attacker can obtain $d_{j1}, d_{j2}, \ldots, d_{jt_2}$ because of the backdoors in $t_2(t_2 < t)$ elliptic curves, he needs to solved $Q_i = d_i P_i$ or $K_i = k_i P_i$ respectively for $d_i$ and $k_i$, which is clearly infeasible because the difficulty of solving the ECDLP, where $i = i1, i2, \ldots, it_1$.

### 4.3 Efficiency Analysis

The computational cost of the proposed MECDSA is mainly determined by addition and multiplication operation in the elliptic curve and inverse operation in the finite prime field. Assume that $t$ is the number of elliptic curves and $l(n_i)$ is the length of the order $n_i$ of the base point $P_i$. Table 1 gives the efficiency analysis of MECDSA and t-ECDSA, including the computational cost and signature length.

Table 1: Efficiency comparison

| Method | Process | Fq Add. | Fq Mut. | Inv. | EC Add. | EC Mul. | Length of Sig. |
|---|---|---|---|---|---|---|---|
| t-ECDSA | Sig. Gen. | t | 2t | t | 0 | t | $2\sum_{i=1}^{t} l(n_i)$ |
|  | Sig. Ver. | 0 | 2t | t | t | 2t |  |
| MECDSA | Sig. Gen. | 2t-1 | 2t | t | 0 | t | $\max(l(n_1), l(n_2), ..., l(n_t)) + t - 1 + \sum_{i=1}^{t} l(n_i)$ |
|  | Sig. Ver. | t-1 | 2t | t | t | 2t |  |

# 5 Conclusion

In order to avoid backdoor in the ECDSA curve of cryptocurrency systems, the paper proposes a multiple elliptic curves digital signature algorithm (MECDSA), which allows the signer to not only choose the number of elliptic curves, but also specify the parameters of each elliptical curve. Our elliptic curves are not limited to the curve P-256 of national institute of standards and technology (NIST)[15], SM2 of Chinese state cryptography administration[16], secp256r111 and secp256k1 in standards for efficient cryptography, and others. We suggest that the systems based on blockchain should choose two elliptic curves considering the contradiction between security and efficiency.

# References

1. Douglas R. Stinson. Cryptography: Theory and Practice. Third Edition, 2006.
2. Ian F. Blake, Gadiel Seroussi, Nigel P. Smart Advances in Elliptic Curve Cryptography. London Mathematical Lecture Note Series, 317, 2005, pp. 3-20.
3. Ujan Mukhopadhyay, Anthony Skjellum, Oluwakemi Hambolu, Jon Oakley, Lu Yu, Richard Brooks. A Brief Survey of Cryptocurrency Systems. 14th Annual Conference on Privacy, Securiy and Trust, 2016, pp. 745-752.
4. Andreas M. Antonopoulos. Mastering Bitcoin. Second Edition, 2017.
5. Dejan Vujičić, Dijana Jagodić, Siniša Ranđić. Blockchain Technology, Bitcoin, and Ethereum: a Brief Overview. 17th International Symposium INFOTEH-JAHORINA, 2018, pp. 1-6.
6. Olivier Alphand, Michele Amoretti, Timothy Claeys. IoTChain: a Block Security Architecture for the Internet of Things. IEEE Wireless Communications and Networking Conference, 2018, pp. 1-6.
7. Emre Yavuz, Ali Kaan Koc, Umut Can Cabuk, Gokhan Dalkilic. Towards Secure E-Voting Using Ethereum Blockchain. 6th International Symposium on Digital Forensic and Security, 2018, PP. 1-7.
8. N. Koblitz. Elliptic Curve Cryptosystems. Mathematics of Computation, 48, 1987, pp. 203-209.
9. A. Mennezes, S. Vanstone. Elliptic Curve Systems. Proposed IEEE P1363 standard, 1995, pp. 1-42.

10. W. J. Caelli, E. P. Dawson, S. A. Rea. PKI, Elliptic Curve Cryptography, and digital signature. Computers & Security, 18, 1999, pp. 47-66.
11. Standards for Efficient Cryptography. SEC 2: Recommended Elliptic Curve Domain Parameters. 2010, available from http://www.secg.org/sec2-v2.pdf.
12. A. J. Menezes, T. Okamoto, S. A. Vanstone. Reducing Elliptic Curve Logarithms to Logarithms in Finite Field. IEEE Transactions on Information Theory, 39, 1993, pp. 1639-1646.
13. I. Biehl, B. Meyer, V. Müller. Differential Fault Attacks on Elliptic Curve Cryptosystems (Extended Abstract). Advances in Cryptology - CRYPTO 2000, LNCS, 1880, 2000, pp. 131-146.
14. J. Blömer, M. Otto, J. P. Seifert. Sign Change Fault Attacks on Elliptic Curve Cryptosystems. Fault Diagnosis and Tolerance in Cryptography - FDTC 2006, LNCS, 4236, 2006, pp. 36-52.
15. National Institute of Standards and Technology (NIST), Digital Signature Standard. In the appendix of FIPS 186-4, available from https://nvlpubs.nist.gov/nistpubs/FIPS/NIST.FIPS.186-4.pdf, July, 2013.
16. State Cryptography Administration, SM2 Elliptic Curve Public Key Cryptography. Aavailable from http://www.sca.gov.cn/sca/xwdt/2010-12/17/content_1002386.shtml, December, 2010.

# Appendix

### 1 Curve P-256 of National Institute of Standards and Technology

For each prime $p$, a pseudo-random curve E: $y^2 = x^3 - 3x + b (mod\, p)$ of prime order n is listed. The following parameters are given:

The prime modulus p:
p=115792089210356248762697446949407573530086143415290314195533631308867097853951

The order n:
n=115792089210356248762697446949407573529996955224135760342422259061068512044369

The coefficient b (satisfying $b^2 c = -27 mod\, p$):
b = 5ac635d8 aa3a93e7 b3ebbd55 769886bc 651d06b0 cc53b0f6 3bce3c3e 27d2604b

The base point P in compressed form is:
P=02 6b17d1f2 e12c4247 f8bce6e5 63a440f2 77037d81 2deb33a0 f4a13945 d898c296
and in uncompressed form is:
P=04 6b17d1f2 e12c4247 f8bce6e5 63a440f2 77037d81 2deb33a0 f4a13945 d898c296 4fe342e2 fe1a7f9b 8ee7eb4a 7c0f9e16 2bce3357 6b315ece cbb64068 37bf51f5

The cofactor h:
h=01

### 2 SM2 of Chinese State Cryptography Administration

The elliptic curve equation is E: $y^2 = x^3 + ax + b$, and the parameters of the equation are as following:

The prime modulus p:

p=fffffffe ffffffff ffffffff ffffffff ffffffff 00000000 ffffffff ffffffff
   The coefficient a:
a= fffffffe ffffffff ffffffff ffffffff ffffffff 00000000 ffffffff fffffffc
   The coefficient b:
b=28e9fa9e 9d9f5e34 4d5a9e4b cf6509a7 f39789f5 15ab8f92 ddbcbd41 4d940e93
   The order n:
n= fffffffe ffffffff ffffffff ffffffff 7203df6b 21c6052b 53bbf409 39d54123
   The base point P in compressed form is:
P=02 32c4ae2c 1f198119 5f990446 6a39c994 8fe30bbf f2660be1 715a4589 334c74c7
and in uncompressed form is:
P=04 32c4ae2c 1f198119 5f990446 6a39c994 8fe30bbf f2660be1 715a4589 334c74c7
bc3736a2 f4f6779c 59bdcee3 6b692153 d0a9877c c62a4740 02df32e5 2139f0a0
   The cofactor h:
h=01

**3 secp256r1 of standards for efficient cryptography**

The verifiably random elliptic curve domain parameters over Fp secp256r1 are specified by the extuple T = (p, a, b, G, n, h) where the finite field Fp is defined by:
p=ffffffff 00000001 00000000 00000000 00000000 ffffffff ffffffff ffffffff
   The curve E: $y^2 = x^3 + ax + b$ over Fp is defined by:
a=ffffffff 00000001 00000000 00000000 00000000 ffffffff ffffffff fffffffc
b=5ac635d8 aa3a93e7 b3ebbd55 769886bc 651d06b0 cc53b0f6 3bce3c3e 7d2604b
   E was chosen verifiably at random as specified in ANSI X9.62 from the seed:
S=c49d3608 86e70493 6a6678e1 139d26b7 819f7e90
   The base point P in compressed form is:
P=03 6b17d1f2 e12c4247 f8bce6e5 63a440f2 77037d81 2deb33a0 f4a13945 d898c296
and in uncompressed form is:
P=04 6b17d1f2 e12c4247 f8bce6e5 63a440f2 77037d81 2deb33a0 f4a13945 d898c296
4fe342e2 fe1a7f9b 8ee7eb4a 7c0f9e16 2bce3357 6b315ece cbb64068 37bf51f5
   The order n is:
n=ffffffff 00000000 ffffffff ffffffff bce6faad a7179e84 f3b9cac2 fc632551
   The cofactor h is:
h=01

**4 secp256k1 of standards for efficient cryptography**

The elliptic curve domain parameters over Fp associated with a Koblitz curve secp256k1 are specified by the sextuple T = (p, a, b, G, n, h), where the finite field Fp is defined by:
p= ffffffff ffffffff ffffffff ffffffff ffffffff ffffffff fffffffe fffffc2f
   The curve E: $y^2 = x^3 + ax + b$ over Fp is defined by:
a=0
b=7
   The base point P in compressed form is:

P=02 79be667e f9dcbbac 55a06295 ce870b07 029bfcdb 2dce28d9 59f2815b 16f81798
and in uncompressed form is:
P=04 79be667e f9dcbbac 55a06295 ce870b07 029bfcdb 2dce28d9 59f2815b 16f81798
483ada77 26a3c465 5da4fbfc 0e1108a8 fd17b448 a6855419 9c47d08f fb10d4b8

Finally the order n of G and the cofactor are:
n=ffffffff ffffffff ffffffff fffffffe baaedce6 af48a03b bfd25e8c d0364141
h=01